\documentclass[%
reprint,
superscriptaddress,
amsmath,amssymb,
aps,
floatfix,
]{revtex4-2}

\usepackage{graphicx}
\usepackage{physics}
\usepackage{dcolumn}

\usepackage{bm}
\usepackage{xcolor}
\usepackage[normalem]{ulem}

\begin{document}
	
	
	\title{The origin of magnetism in a supposedly nonmagnetic osmium oxide}
	
	\author{S. Agrestini}
	\affiliation{Diamond Light Source, Harwell Campus, Didcot OX11 0DE, United Kingdom}
	
	\author{F. Borgatti}
	\affiliation{Istituto per lo Studio dei Materiali Nanostrutturati, Consiglio Nazionale delle Ricerche (CNR-ISMN), Via P. Gobetti 101, I-40129 Bologna, Italy}
	
	\author{P. Florio}
	\affiliation{Dipartimento di Fisica, Politecnico di Milano, Piazza Leonardo da Vinci 32, I-20133 Milano, Italy}
	
	\author{J. Frassineti}
	\affiliation{Dipartimento di Fisica e Astronomia, Alma Mater Studiorum - Università di Bologna, Viale C. Berti Pichat 6/2, I-40127 Bologna, Italy}
	
	\author{D. Fiore Mosca}
	\address{Centre de Physique Th\'eorique, Ecole Polytechnique, CNRS, Institut Polytechnique de Paris, 91128 Palaiseau Cedex, France}
	
	\author{Q. Faure}
	\affiliation{European Synchrotron Radiation Source, 71 Avenue des Martyrs, F-38000 Grenoble, France}
	
	\author{B. Detlefs}
	\affiliation{European Synchrotron Radiation Source, 71 Avenue des Martyrs, F-38000 Grenoble, France}
	
	\author{C. J. Sahle}
	\affiliation{European Synchrotron Radiation Source, 71 Avenue des Martyrs, F-38000 Grenoble, France}
	
	\author{S. Francoual}
	\affiliation{Deutsches Elektronen-Synchrotron DESY, Notkestr. 85, D-22607 Hamburg, Germany}
	
	\author{J. Choi}
	\affiliation{Diamond Light Source, Harwell Campus, Didcot OX11 0DE, United Kingdom}
	
	\author{M. Garcia-Fernandez}
	\affiliation{Diamond Light Source, Harwell Campus, Didcot OX11 0DE, United Kingdom}
	
	\author{K.-J. Zhou}
	\affiliation{Diamond Light Source, Harwell Campus, Didcot OX11 0DE, United Kingdom}
	
	\author{V. F. Mitrovic}
	\affiliation{Department of Physics, Brown University, Providence, Rhode Island 02912, USA}
	
	\author{P. M. Woodward}
	\affiliation{Department of Chemistry and Biochemistry, The Ohio State University, Columbus, Ohio 43210, USA}
	
	\author{G. Ghiringhelli}
	\affiliation{Dipartimento di Fisica, Politecnico di Milano, Piazza Leonardo da Vinci 32, I-20133 Milano, Italy}
	
	\author{C. Franchini}
	\affiliation{Dipartimento di Fisica e Astronomia, Alma Mater Studiorum - Università di Bologna, Viale C. Berti Pichat 6/2, I-40127 Bologna, Italy}
	
	\author{F. Boscherini}
	\affiliation{Dipartimento di Fisica e Astronomia, Alma Mater Studiorum - Università di Bologna, Viale C. Berti Pichat 6/2, I-40127 Bologna, Italy}
	
	\author{S. Sanna}
	\affiliation{Dipartimento di Fisica e Astronomia, Alma Mater Studiorum - Università di Bologna, Viale C. Berti Pichat 6/2, I-40127 Bologna, Italy}
	
	\author{M. Moretti Sala}
	\affiliation{Dipartimento di Fisica, Politecnico di Milano, Piazza Leonardo da Vinci 32, I-20133 Milano, Italy}
	
	\date{\today}
	
	\begin{abstract}
		
		A supposedly nonmagnetic 5d$^1$ double perosvkite oxide is investigated by a combination of spectroscopic and theoretical methods, namely resonant inelastic X-ray scattering, X-ray absorption spectroscopy, magnetic circular dichroism, and multiplet ligand field calculations. We found that the large spin-orbit coupling admixes the 5d $t_{2g}$ and $e_g$ orbitals, covalency raises the 5d population well above the nominal value, and the local symmetry is lower than $O_h$. The obtained electronic interactions account for the finite magnetic moment of Os in this compound and, in general, of 5d$^1$ ions. Our results provide direct evidence of elusive Jahn-Teller distortions, hinting at a strong electron-lattice coupling.
		
	\end{abstract}
	
	\maketitle
	
	The field of spin-orbit coupled Mott insulators has been thriving since the notion that spin-orbit interaction can induce unexpected behaviors in solids was put forward more than a decade ago~\cite{BJKim2008,BJKim2009}. In particular, 5d transition metal compounds provide a platform to explore the rich physics stemming from the strong spin-orbit interaction and the relatively large spatial extension of atomic orbitals: the former admixes orbitals of different symmetry and gives rise to entangled spin and orbital magnetic moments; the latter attenuates electronic correlations and promotes covalency. Their combination produces an intricate network of interactions, comparable in strength, that connects all electronic degrees of freedom~\cite{Witczak-Krempa2014}. So far, much of the work has concentrated on the study of entangled spin-orbital states in systems with different band fillings and on multiple bond geometries motivated by the number of exotic electronic and magnetic interactions they give rise to ~\cite{Jackeli2009,Rau2016,Winter2017,Cao2018}. However, little attention has been paid to other, equally intriguing effects, such as the coupling of electronic degrees of freedom to the lattice~\cite{Liu2019,Streltsov2020,Streltsov2022}. A number of theoretical predictions suggests that electron-lattice coupling is relevant for spin-orbit Mott insulators, particularly in cubic 5d$^1$ systems: indeed, in the limit of very large crystal field, high energy e$_g$ states do not mix in and otherwise degenerate t$_{2g}$ states are split by spin-orbit coupling into an empty $j_\mathrm{eff}=1/2$ Kramers doublet and a singly occupied $j_\mathrm{eff}=3/2$ quartet, consisting of \textit{two} Kramers doublets; the extra double degeneracy leads to Jahn-Teller (JT) instabilities and generates strong quantum effects~\cite{Streltsov2022}. In the specific case of double perovskite Ba$_2$NaOsO$_6$~\cite{Stitzer2002}, it was theoretically predicted that electron-lattice coupling results in entangled spin-orbital-lattice states~\cite{Xu2016} featuring multiferroic effects~\cite{Iwahara2018}.
	
	The electronic structure of Ba$_2$NaOsO$_6$ is mostly determined by molecular orbitals of weakly interacting, allegedly undistorted OsO$_6$ octahedra, the Os ions having nominal Os$^{7+}$ (5d$^1$) and local $O_h$ site symmetry~\cite{Erickson2007}. DFT calculations show that electronic correlation cannot sustain an insulating behaviour by itself unless assisted by a sizeable spin orbit coupling, thus confirming the spin-orbit coupled Mott
	insulating nature of Ba$_2$NaOsO$_6$~\cite{Xiang2007,Lee2007,Cong2019}. Noticeably, despite the magnetic moment for the single 5d$^1$ electron occupying the $j_\mathrm{eff}=3/2$ states should be zero by virtue of the exact cancellation of its spin and orbital angular momenta~\cite{Kotani1949,Abragam1970,Mabbs1973},
	Ba$_2$NaOsO$_6$ features an effective magnetic moment $\mu_\mathrm{eff}\approx 0.6$ $\mu_B$ in the paramagnetic phase \cite{Erickson2007, Cong2023} and a net moment of 0.2 $\mu_B$ in the long-range ordered canted
	antiferromagnetic (c-AFM) phase below $T_N$ = 6.8 K ~\cite{Stitzer2002,Erickson2007}. Recent DFT calculations show that this exotic magnetic order is understood and originates from the anomalous Dzyaloshinskii-Moriya interaction generated by JT distortions~\cite{Mosca2021}. But, why is Ba$_2$NaOsO$_6$ magnetic at all to start with? More DFT and \textit{ab initio} calculations suggest that the sizable magnetic moment found experimentally could be explained by strong Os 5d--O 2p hybridization~\cite{Lee2007,Xu2016,Ahn2017} and/or by JT effect~\cite{Xiang2007,Xu2016,Iwahara2018},
	but are not backed up by experiments: NMR~\cite{Lu2017,Cong2023}, specific heat and magnetization~\cite{Willa2019} data evidence a broken symmetry phase up to only a few Kelvin above the magnetic transition, while X-ray~\cite{Erickson2007} and neutron~\cite{Barbosa2022} diffraction, and EXAFS experiments~\cite{Kesavan2020} show that the space group symmetry of Ba$_2$NaOsO$_6$ ($Fm\bar{3}m$) is cubic at higher temperatures, with no evidence of JT distortions. The sole experimental hint about a possible room temperature splitting of the $j_\mathrm{eff}=3/2$ quartet into two Kramers doublets comes from specific heat measurements~\cite{Erickson2007}. The mismatch between theoretical predictions and experimental evidences calls for further investigations.
	
	In this letter, we address the question related to the origin of magnetism in Ba$_2$NaOsO$_6$ by spectroscopic methods sensitive to the bulk electronic structure and magnetic state of the Os ions, namely resonant inelastic X-ray scattering (RIXS), X-ray absorption spectroscopy (XAS) and magnetic circular dichroism (XMCD). The analysis of our experimental results, supported by multiplet ligand field theory (MLFT) calculations, shows that covalency increases the occupancy of the 5d states well beyond the nominal d$^1$ configuration, and provides solid experimental evidence of a JT distortion in this system already at room temperature. In order to reconcile our and previous results, a dynamic mechanism for the JT distortion is proposed.
	
	\begin{figure}
		\centering
		\includegraphics[width=0.9\columnwidth]{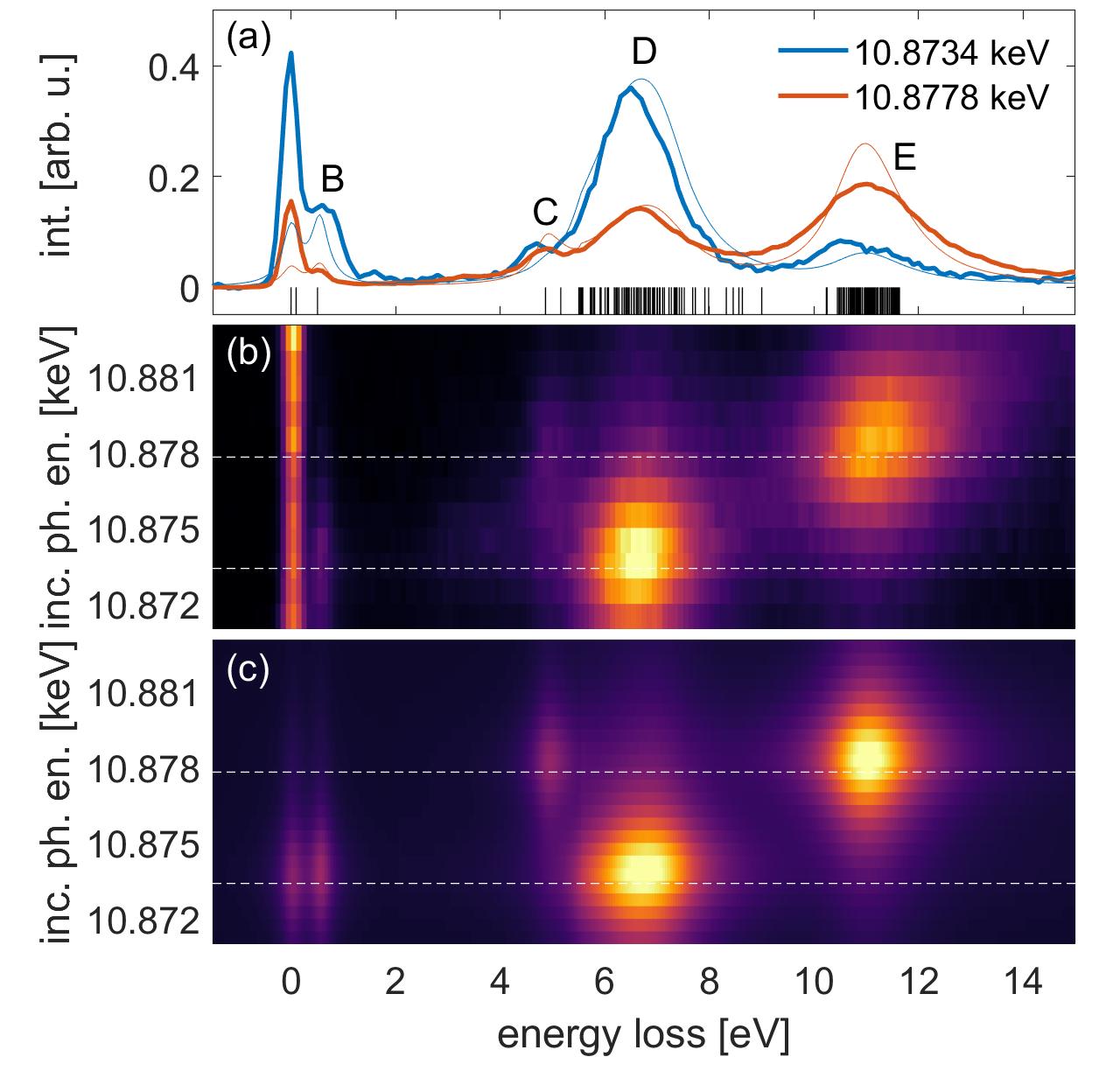}
		\caption{(a) Experimental (thick) and simulated (thin lines) Os L$_3$ edge RIXS spectra of powder Ba$_2$NaOsO$_6$ at 300 K measured at selected incident photon energies, as extracted from the RIXS map of panels (b) and (c), respectively, at incident photon energies corresponding to the white dashed lines. Calculations of the excitation energies, vertical lines in the bottom of panel (a), and intensities were performed by MLFT, as explained in the text.}
		\label{fig:1}
	\end{figure}
	The synthesis of powder samples of Ba$_2$NaOsO$_6$ is described in Ref.~\cite{Kesavan2020}, while single crystals are the same as in Ref.~\cite{Lu2017}. Os L$_3$ edge RIXS measurements were collected at beamline ID20 ~\cite{MorettiSala2018} at the ESRF, O K edge RIXS measurements at I21 beamline~\cite{Zhou2022} at the DLS, and Os L$_{2,3}$ edge XAS/XMCD measurements at beamline P09~\cite{Strempfer2013,Strempfer2016} at PETRA III at DESY. MLFT calculations on a OsO$_6$ cluster were carried out with Quanty~\cite{Haverkort2012,Lu2014,Haverkort2014}, which explicitly takes into account the full atomic multiplet, including spin-orbit coupling, crystal field splitting acting on the metal and on the ligand ions, as well as their hybridization.
	
\begin{figure}
		\centering
		\includegraphics[width=0.9\columnwidth]{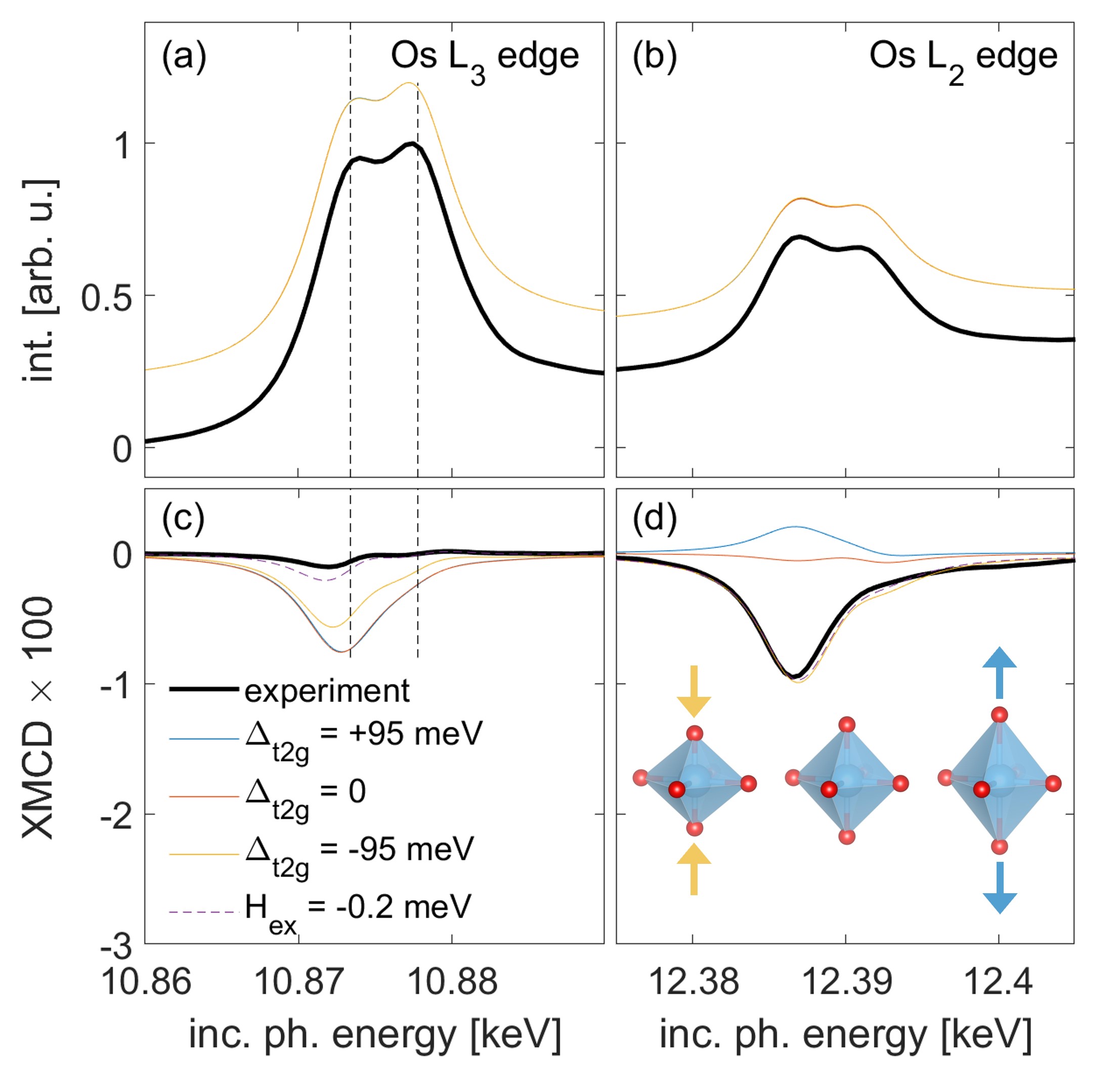}
		\caption{Os L$_3$ (a,c) and L$_2$ (b,d) edge XAS and XMCD curves at 3.5 K. The thick black and thin coloured lines correspond to experimental and simulated curves, respectively. The vertical dashed lines in panels (a) and (c) correspond to the incident photon energies used to measure ad simulate the RIXS spectra of Fig.~\ref{fig:1}(a). The inset of panel (d) shows compressed, undistorted and elongated OsO$_6$ octahedra, which refer to the simulated yellow, red and blue curves, respectively.}    \label{fig:3}
	\end{figure}

	Fig.~\ref{fig:1}(b) shows the Os L$_3$ edge RIXS map of powder Ba$_2$NaOsO$_6$ at 300 K, measured with a low energy-resolution, high throughput setup over a large range of incident and emitted photon energies in order to obtain a broad overview of the RIXS response in this system. The RIXS map is very rich and, besides the elastic line at zero energy loss, a number of Raman-like excitations are clearly visible, resonating at incident photon energies roughly corresponding to the two peaks in the XAS profile [black curve in Fig.~\ref{fig:3}(a)]. The main excitations (labelled B, C, D and E) are better appreciated in the two spectra of Fig.~\ref{fig:1}(a), obtained by cutting the RIXS map at constant incident photon energies. We note that: i) peaks B (0.5 eV) and C (4.9 eV) are relatively weak, compared to peaks D (6.5 eV) and E (11.0 eV). However, their energy separation (4.4 and 4.5 eV, respectively) is very similar; ii) peaks B and D resonate at 10.8734 keV, while peaks C and E resonate at 10.8778 keV, i.e. 4.4 eV higher in energy; iii) the energy separation between peaks D and B and between peaks C and E (6.0 and 6.1 eV, respectively) is also very similar. The natural interpretation of all these features is in terms of ligand-field transitions; in particular, we assign peaks B and C to transitions within the Os 5d states, involving t$_{2g}$ and e$_g$ states, respectively, while peaks D and E to transitions from O 2p to Os 5d states, again involving t$_{2g}$ and e$_g$ states, respectively. We term the former $d$-$d$ and the latter \textit{charge-transfer} (CT) excitations and note that, unlike other compounds, e.g., cuprates~\cite{MorettiSala2011}, the fact that CT overwhelm $d$-$d$ transitions suggests a strong covalent nature of the Os--O bond in Ba$_2$NaOsO$_6$.
	
	Simulated RIXS spectra for optimal values of the parameters~\cite{calc_par} are shown in Fig.~\ref{fig:1}(c) after convolution with the instrumental energy resolution for an appropriate comparison with the experiments. The simulations reproduce the experimental data remarkably well, including the energies of the various excitations, their relative intensities and the differences in their resonant behavior. In particular, MLFT calculations confirm the assignment above: in analogy to K$_2$TaCl$_6$ and Rb$_2$TaCl$_6$~\cite{Ishikawa2019}, peak B is associated to the $j_\mathrm{eff}=3/2$-to-$j_\mathrm{eff}=1/2$ transition, whose energy is primarily set by the spin-orbit coupling constant, while peak C to t$_{2g}$-to-e$_g$ transitions and its energy corresponds to the effective splitting of the t$_{2g}$ and e$_g$ orbitals (10Dq$^\mathrm{eff}=4.9$~eV), resulting from both the ionic and covalent nature of the bonding. Information on the degree of covalency of the system is best extracted from CT transitions (peaks D and E); their energies and intensities are mostly influenced by the strength of the Os 5d--O 2p hybridization and the CT energy: we find a charge-transfer energy $\Delta=-4$~eV, classifying Ba$_2$NaOsO$_6$ as a negative charge-transfer system~\cite{Mizokawa1991,khomskii2001}. Remarkably, in order to reproduce the relative intensities of $d$-$d$ and CT excitations, it has been necessary to expand the MLFT ground state wave function to include configurations with up to four ligand holes, i.e.,
	\begin{equation}\label{eq:ratio}
		\ket{\Psi} = \sum_{i=1}^5\alpha_i\ket{5\mathrm{d}^i\underline{\mathrm{L}}^{i-1}},
	\end{equation}
	where $\underline{\mathrm{L}}$ denotes a ligand hole and $\sum^{5}_{i=1}\left|\alpha_i\right|^2=1$. We find that the nominal 5d$^1$ configuration is negligible (0.01\%), while the 5d$^{2}\underline{\mathrm{L}}$ (8.5\%), 5d$^{3}\underline{\mathrm{L}}^2$ (29.0\%), 5d$^{4}\underline{\mathrm{L}}^3$ (41.6\%), and 5d$^{5}\underline{\mathrm{L}}^4$ (20.9\%) configurations are dominant. Ultimately, the number of electrons effectively populating the Os 5d states is $n_d=\sum_{i=1}^5 i\left|\alpha_i\right|^2 =3.75$, meaning that almost three electrons are transferred from the O ligands to the central Os ion. This significant transfer of charge is consistent with DFT calculations, which report an even higher population of the Os 5d states, i.e., $n_d \approx 5-6$~\cite{Cong2019}.
	
	\begin{figure}
		\centering
		\includegraphics[width=0.9\columnwidth]{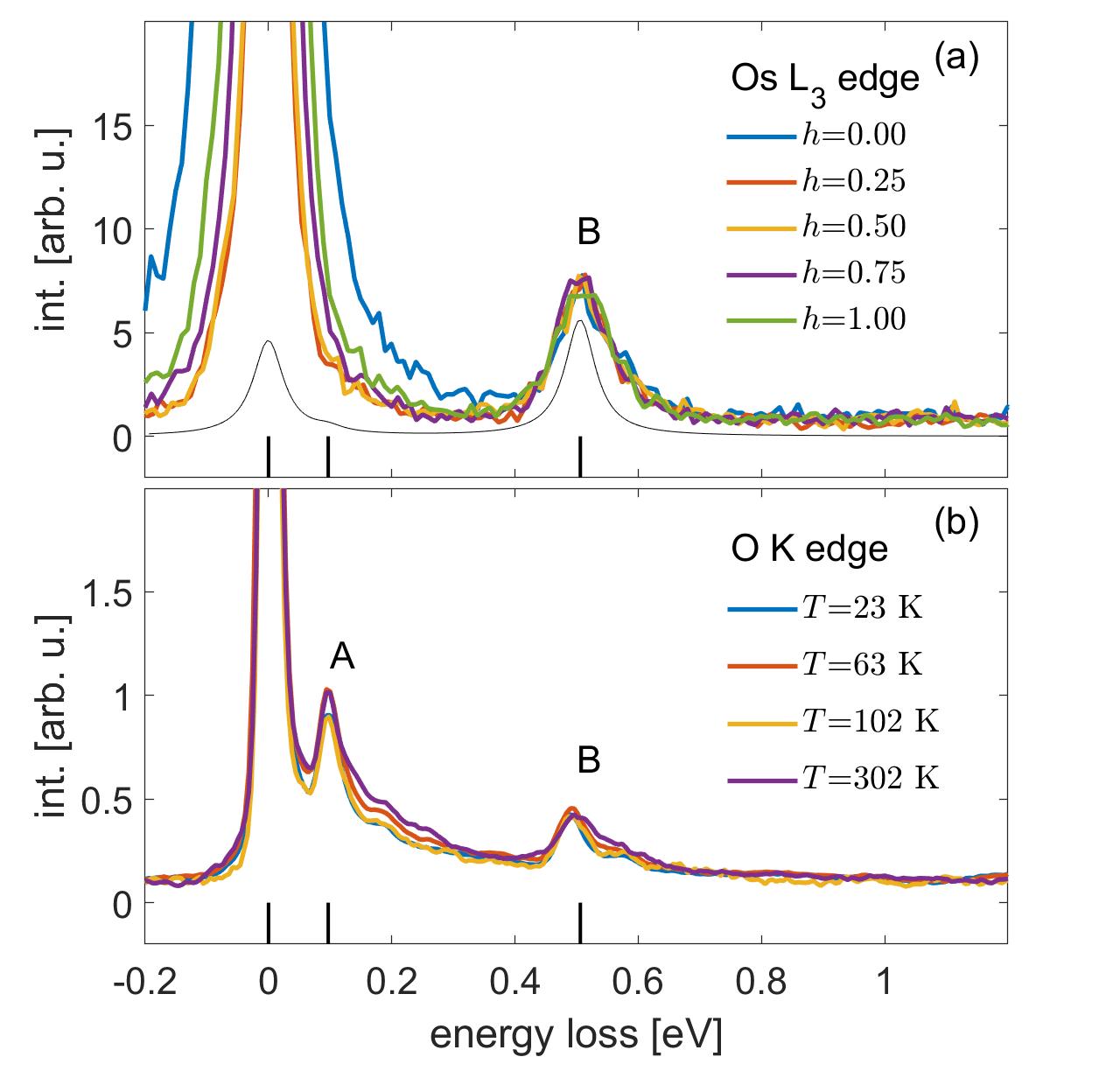}
		\caption{(a) Os L$_3$ edge RIXS spectra of Ba$_2$NaOsO$_6$ single crystal at 300 K as a function of the transferred momentum along the (10+$h$,0,0) r.l.u. direction in reciprocal space. The thin black line shows the simulated RIXS spectrum in the experimental geometry corresponding to (10.5,0,0) r.l.u. momentum transfer. (b) O K edge RIXS spectra of powder Ba$_2$NaOsO$_6$ as a function of temperature. Calculations of the excitation energies, vertical lines in panel (a), were performed by MLFT, as explained in the text.}
		\label{fig:2}
	\end{figure}

	Fig.~\ref{fig:2}(a) shows the Os L$_3$ edge RIXS spectra of a Ba$_2$NaOsO$_6$ single crystal at 300 K, measured with a low throughput, high energy-resolution setup as a function of momentum transfer in a small energy range close to the elastic line at 10.8734 keV incident photon energy. Both the elastic line and peak B are slightly asymmetric, and peak B shows no momentum transfer dependence. The latter observation is consistent with the assumption that the electronic structure of Ba$_2$NaOsO$_6$ is mostly determined by molecular orbitals of weakly interacting, OsO$_6$ octahedra~\cite{Erickson2007}, thus justifying the use of MLFT calculations to perform a quantitative analysis of the experimental RIXS spectra and extract significant physical parameters.
	
	So far, our experimental results support the notion that the electronic structure of Ba$_2$NaOsO$_6$ is mostly determined by the molecular orbitals of undistorted OsO$_6$ octahedra. The main challenge to this paradigm comes from O K edge RIXS and Os L$_{2,3}$ edge XMCD measurements. Fig.~\ref{fig:2}(b) shows O K edge RIXS spectra of a Ba$_2$NaOsO$_6$ single crystal, collected at various temperatures in the range 21--302 K. Peak B at 0.5 eV is visible and accompanied by a weak satellite at higher energy; in addition, we find a sharp excitation at 0.095 eV (labelled A), which shows neither systematic temperature nor momentum transfer dependence (see \cite{SuppMat}), therefore excluding its magnetic or phononic origin. Peak A lies on top of a broad distribution of spectral weight extending from the elastic line up to approximately 0.4 eV, which can possibly arise from multi-phonon excitations resonating at approximately 1 eV higher incident photon energy with respect to peak A (see \cite{SuppMat}). Its most plausible interpretation is in terms of a $d$-$d$ excitation, and implies the breaking of the octahedral ($O_h$) site symmetry and the lifting of the $j_\mathrm{eff}=3/2$ degeneracy into two doublets, between which RIXS transitions can occur.  
	
	In order to further assess the nature of peak A and test the hypothesis of the breaking of the Os site symmetry to lower than octahedral, e.g., tetragonal ($D_{4h}$), we performed XMCD measurements both in the c-AFM and paramagnetic phase.
	Os L$_{2,3}$ edge XAS and XMCD data measured at 3.5 K in a 5 T magnetic field are reported as continuous black lines in Fig.~\ref{fig:3}(a) and (b). The spectra at 40 K (not shown) are qualitatively similar, but with reduced intensity. We note two unusual features: i) the XMCD signal is negative at both L$_3$ and L$_2$ edges; ii) almost all of the XMCD intensity is at the L$_2$ absorption edge.
	We then performed MLFT calculations with the same parameters used to simulate the Os L$_3$ edge RIXS data above, but varying the tetragonal crystal field $\Delta_{t2g}$. In case of $O_h$ symmetry [$\Delta_{t2g}=0$, red curves in Fig.~\ref{fig:3}(c) and (d)], the simulation predicts a strong XMCD signal at the Os L$_3$ edge and a weak one at L$_2$, in stark contrast to the experiments. Only the introduction of a negative tetragonal crystal-field changes qualitatively the shape of the XMCD curves and considerably improves the agreement between simulated and experimental spectra (see \cite{SuppMat}): the best agreement to the L$_2$ edge XMCD signal is obtained for $\Delta_{t2g}=-0.095$~eV, which nicely matches the energy of peak A in O K edge RIXS [Fig.~\ref{fig:2}(b)]. It corresponds to a tetragonal compressive distortion of the OsO$_6$ octahedra along the $z$-axis, which, in absence of spin-orbit coupling, would lower the energy of the $xy$ orbital to below that of the $yz$ and $zx$ orbitals. It is important to note that the simulated intensity of peak A for the proposed tetragonal distortion is too weak to be detected by Os L$_3$ edge RIXS [thin black line in Fig.~\ref{fig:2}(a)].
	
\begin{figure}
		\centering
		\includegraphics[width=0.9\columnwidth]{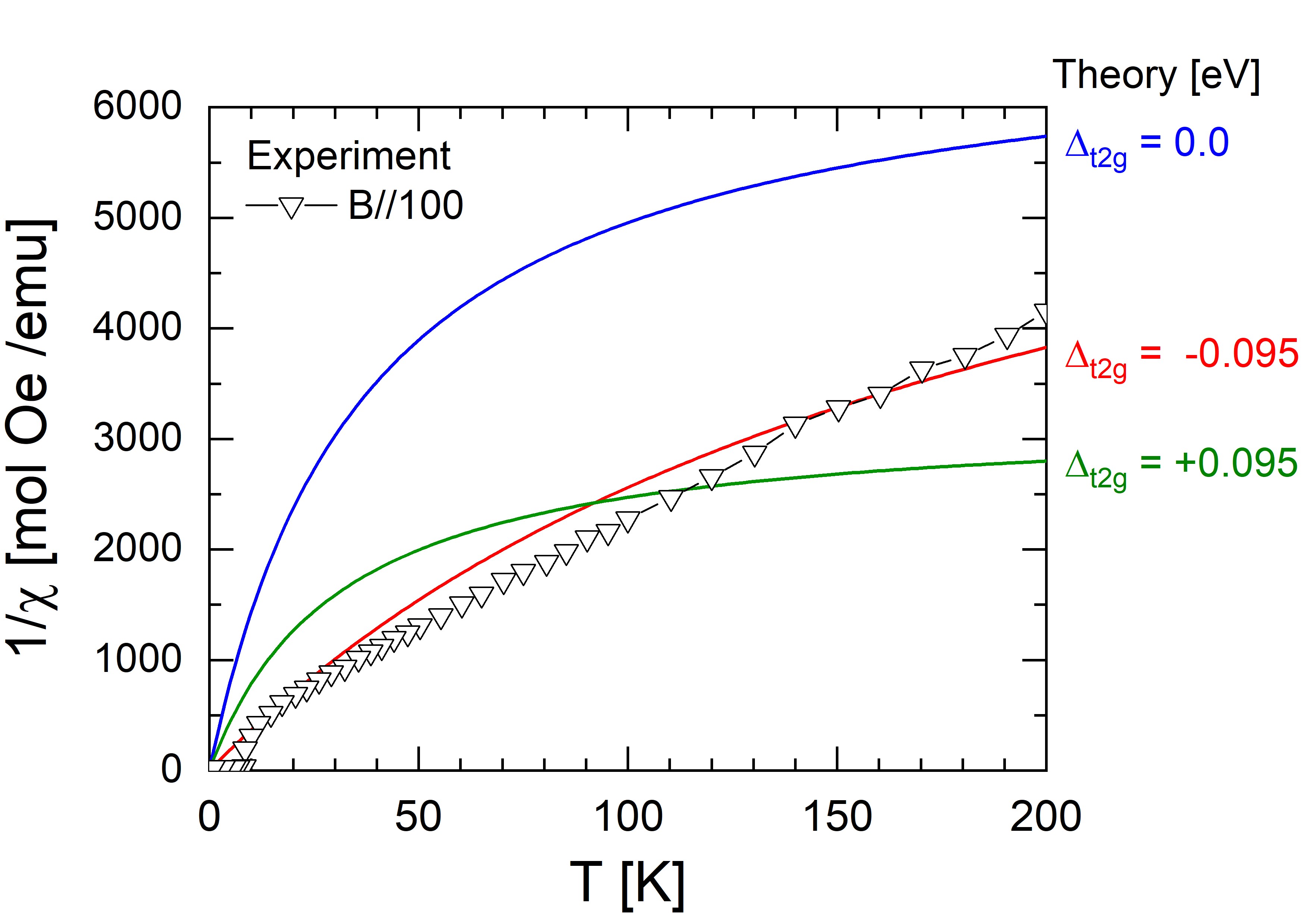}
		\caption{Experimental (open triangles, from Ref. \cite{Erickson2007}) and simulated (continuous lines) inverse magnetic susceptibility of Ba$_2$NaCaOsO$_6$ as a function of temperature.}
		\label{fig:4}
	\end{figure}

	In essence, O K edge RIXS and Os L$_{2,3}$ edge XMCD measurements provide compelling evidence that Ba$_2$NaOsO$_6$ is JT active at all temperatures. On the one hand, our results agree with heat capacity measurements (JT distortions splits the $j_\mathrm{eff} = 3/2$ states into two Kramers doublets, hence the integrated entropy through the magnetic phase transition of $R\ln 2$)~\cite{Erickson2007}, quantitatively reproduce the temperature dependence of the magnetic susceptibility in the paramagnetic phase (see Fig. \ref{fig:4}) and support theoretical calculations for Ba$_2$NaOsO$_6$~\cite{Xu2016,Iwahara2018}; on the other hand, they seemingly contradict X-ray diffraction~\cite{Erickson2007}, NMR~\cite{Lu2017,Liu2018} and EXAFS~\cite{Kesavan2020} studies, although a tetragonal distortion could not be ruled out by neutron diffraction~\cite{Barbosa2022}. In order to reconcile our and previous experiments, we speculate that Ba$_2$NaOsO$_6$ is subject to a dynamic JT effect: the system resonates between equivalent minima of the adiabatic potential energy surface (corresponding to different directions of the distortions), such that the average crystal structure remains undistorted, explaining why such distortions could not be observed directly by structural probes~\cite{Bersuker2006}. Moreover, its dynamical nature makes the JT distortion invisible to slow spectroscopic probes~\cite{Bersuker2006}, such as NMR.
	
	Next, we calculate the expectation values of the magnetic moment ($\mu$) at T = 20 K (paramagnetic phase) within MLFT (see \cite{SuppMat}) to clarify the role of the various interactions in determining the magnetic properties of Ba$_2$NaOsO$_6$. In $O_h$ symmetry we find that for 10Dq$^\mathrm{eff}=\infty$ the $L_x/2S_x$ ratio is -1.00, meaning that the spin and orbital angular momenta are equal in size and have opposite sign, such that none of the $j_\mathrm{eff} = 3/2$ states are magnetic ($\mu_x=0$); but for a realistic, although large value of 10Dq$^\mathrm{eff}$ (4.9 eV) a strong spin-orbit coupling mixes the $j_\mathrm{eff} = 3/2$ ($m_j=\pm 3/2$) and the e$_g$ inducing a finite magnetic moment ($\mu_x=0.007$ $\mu_B$). Covalency causes a partial suppression of the orbital angular moment, so that the magnetic moment reduces to $\mu_x=0.004$ $\mu_B$. Lowering the symmetry from $O_h$ to $D_{4h}$ due to a finite value of $\Delta_{t2g}$ (-0.095~eV) has an important effect: the $xy$ plane become an easy plane of magnetization and the magnetic moment is increased with respect to the $O_h$ case to $\mu_x = 0.014$ $\mu_B$ in presence of covalency. 
	
	The simulation of the small XMCD signal at the Os L$_3$ edge remains unsatisfactory at this stage, but can be improved, as explained in the following. The tetragonal compressive distortion of the OsO$_6$ octahedra [sketched in the inset of Fig.~\ref{fig:3}(d)] favors the alignment of the magnetic moments in the $xy$ plane, consistent with the magnetic structure proposed in Ref.~\cite{Lu2017}. It turns out that in this case the XMCD signal at the Os L$_3$ edge is very sensitive to the presence of magnetic couplings between the Os ions; in particular, a fairly good agreement can be obtained by introducing an exchange field H$_{ex}$ that mimics at a mean field level the weak antiferromagnetic coupling between nearest-neighbor magnetic moments. The best simulation is obtained for -0.2~meV [dashed purple curves in Fig.~\ref{fig:3}(c) and (d)], which roughly compares to the ordering temperature ($k_BT_N\approx 0.6$ meV)~\cite{Erickson2007}, thus strengthening the intuition that the reduced Os L$_3$ edge XMCD intensity originates from Os-Os magnetic interactions.
	
	In summary, our spectroscopic investigations provide a quantitative estimate of the relevant electronic interactions in Ba$_2$NaOsO$_6$; all together, they account for its finite magnetic moment and solve the puzzle concerning the origin of magnetism in this material. In particular, spin-orbit coupling and dynamic JT effects, which RIXS is sensitive to~\cite{Iwahara2023}, entangle the spin, orbit and lattice degrees of freedom and could give rise to exotic magnetic (and vibronic)
	orders~\cite{Iwahara2023bis}. Beyond the specific case of Ba$_2$NaOsO$_6$, our results are relevant for several (anti)ferromagnetic 4d$^1$ \cite{Aharen2010,Ishikawa2021} and 5d$^1$ \cite{Steele2011,Yamamura2006,Tehrani2021,Ishikawa2021,Ishikawa2019} double perovskites, as well as other systems with exotic magnetic ground states\cite{Cussen2006,Wiebe2002,Wiebe2003}.
	
	\begin{acknowledgments}
		We acknowledge Ian Fisher for providing the single crystalline samples. We acknowledge the ESRF (Grenoble, France), DLS (Didcot, United Kingdom) and DESY (Hamburg, Germany), a member of the Helmholtz Association HGF, for provision of synchrotron radiation facilities. Parts of this research were carried out at beamline ID20 at ESRF under proposals HC-4277 and HC-4912. We would like to thank F. Gerbon for assistance and support in using the beamline. Parts of this research were carried out at beamline P09 at PETRA III under proposal I-20200367EC. We would like to thank Dr. J. R. Linares Mardegan and Dr. T. Pohlmann for their help with the measurements, and Dr. Olaf Leupold for assistance in cooling down and operating the 6T/2T/2T magnet at P09. The 6T/2T/2T magnet used for the XMCD measurements was funded in part by the BMBF grant No. 05K2013 from the German Federal Ministry of Education and Research. The work here presented is partly funded by the European Union – Next Generation EU - ``PNRR - M4C2, investimento 1.1 - Fondo PRIN 2022'' - ``Superlattices of relativistic oxides''  (ID 2022L28H97, CUP D53D23002260006) and ``Spin-charge-lattice coupling in relativistic Mott insulators'' (ID 202243JHMW, CUP J53D23001350006).
	\end{acknowledgments}
	
	\bibliography{biblio}

\begin{thebibliography}{52}
\bibitem [{cal()}]{calc_par}%
  \BibitemOpen
  \href@noop {} {}\bibinfo {note} {OsO$_6$ cluster parameters [eV]: U$_{dd}$ =
  1.0, U$_{pd}$ = 3.0, charge transfer energy $\Delta_{CT}=-4.0$, spin-orbit
  coupling $\lambda=0.41$, ionic crystal field 10Dq$^\mathrm{ion}=3.3$,
  $\Delta_{t_{2g}}^\mathrm{ion}=-0.25$ and $\Delta_{e_g}^\mathrm{ion}=-0.5$,
  hybridization $V(e_g)=6.2$ and $V(t_{2g})=3.5$, ligand crystal field
  10Dq$^\mathrm{lig}=0.95$. Slater integrals were reduced to 70~\% of the
  Hartree-Fock values. The effective cubic crystal field is
  10Dq$^\mathrm{eff}=4.9$~eV. For all simulations we have considered the
  thermal population of the electronic state using the Boltzmann
  distribution.}\BibitemShut {Stop}%
\end{thebibliography}
	
\end{document}